\documentclass[aps,prl,superscriptaddress,twocolumn,floatfix]{revtex4-2}
\usepackage{setspace}
\usepackage[T1]{fontenc}
\usepackage[latin9]{inputenc}
\usepackage{amstext}
\usepackage{amsmath}
\usepackage{amsfonts}
\usepackage{graphicx}
\usepackage{graphics}
\usepackage{subfigure}
\usepackage{esint}
\usepackage{color}
\usepackage{physics}
\usepackage[normalem]{ulem}
\usepackage[colorlinks, linkcolor=blue, citecolor=blue,hyperindex,bookmarks=false,pdfstartview=FitH]{hyperref}
\usepackage{lineno}
\makeatletter

\@ifundefined{textcolor}{}
{%
 \definecolor{BLACK}{gray}{0}
\definecolor{WHITE}{gray}{1}
 \definecolor{RED}{rgb}{1,0,0}
 \definecolor{GREEN}{rgb}{0,1,0}
 \definecolor{BLUE}{rgb}{0,0,1}
 \definecolor{CYAN}{cmyk}{1,0,0,0}
 \definecolor{MAGENTA}{cmyk}{0,1,0,0}
 \definecolor{YELLOW}{cmyk}{0,0,1,0}
}
\makeatother

\makeatother

\begin{document}

\title{Local Counterdiabatic Driving for Jaynes-Cummings Lattices}

\author{Anuvetha Govindarajan}
\affiliation{School of Natural Sciences, University of California, Merced, California 95343, USA}

\author{Lin Tian}
\email{ltian@ucmerced.edu}
\affiliation{School of Natural Sciences, University of California, Merced, California 95343, USA}

\begin{abstract}
Jaynes-Cummings (JC) lattices can be constructed by connecting quantum two-level systems with cavities and have been widely studied for polariton many-body states and multipartite entanglement. Although adiabatic evolution has been studied for the generation of many-body states in this system, its reliance on long timescales can lead to serious decoherence. Here we present a scheme that utilizes local counterdiabatic (CD) driving to provide fast and high-fidelity state preparation in JC lattices. The exact CD Hamiltonian for this system contains nonlocal couplings between qubits and cavities at different and distant sites, which causes a challenge in the implementation. Leveraging the symmetries of the eigenstates under both periodic and open boundary conditions, we derive a local CD Hamiltonian that generates the same dynamics as the exact CD Hamiltonian and our numerical simulations confirm this result. We also show that a multipartite W-state can be prepared with high fidelity using this method. The implementation and decoherence of this scheme with superconducting quantum devices are also discussed.   
\end{abstract}
\maketitle

\emph{Introduction}. The current state-of-art in quantum devices offers remarkable controllability over both individual components and their interactions~\cite{squbit_rev}. The coupling between qubits and cavities as well as the cavity-cavity coupling can be manipulated and engineered through various approaches~\cite{SiddiqiPRL2021, CampbellLaHayePRApplied2023, Chen:2014, FYanPRApplied2018, SandbergAPL2008}. These developments have enabled the generation of complex quantum states, including multipartite entangled states and correlated many-body states, and paved the way for exploring novel quantum dynamics and quantum information applications. One of the models that have been intensively studied owing to the advances in quantum devices is the Jaynes-Cummings (JC) model, which is a cornerstone of quantum optics and describes a two-level system (qubit) coupled to the electromagnetic field of a cavity mode~\cite{Larson2022Review, BlaisRMP2021cQED}. JC lattices consisting of arrays of JC models can demonstrate rich physical phenomena such as quantum phase transitions, making them a valuable platform for quantum simulation and computation~\cite{Noh2017Review, 2012HouckNP_JCQS, BWLi2022Ion, HouckPRX2017, Sala2015Nanophotonics, Seo2015:1, Xue2017, TianPRL2011}. 

Preparing desired quantum states in JC lattices is essential for studying these phenomena. One approach for preparing such quantum states is via adiabatic evolution, which involves a slow transformation of a quantum system from an initial state to a desired final state governed by a time-varying Hamiltonian~\cite{Albash2018, FarhiScience2001}. However, the requirement for slow evolution in the adiabatic approach can be impractical in noisy quantum devices due to the extended timescales, which increase the susceptibility to environmental decoherence. Meanwhile, fast evolution (short timescales) often induces unwanted diabatic transitions to excited states, which can reduce the fidelity of the generated states.  
To mitigate the decoherence issues in the adiabatic approach, shortcuts to adiabaticity (STA) have been developed, which result in the generation of desired final states in a much shorter evolution time and have been demonstrated in experiments~\cite{Guery-Odelin2019, Kolodrubetz2017, UnanyanOptCommon1997, EmmanouilidouPRL2000, Wang2018, Du2016, Zhou2017}. One of the STA methods is the counterdiabatic (CD) driving approach that utilizes a CD Hamiltonian to eliminate diabatic transitions 
~\cite{Berry2009, Demirplak2008, Chen:PhysRevLett:2010:105, delCampoPRL2012, Damski2014}. Despite its advantages, the primary problem of STA is that it almost always requires nonlocal or multipartite interactions, which are challenging to implement in practical devices. Recently, there have been growing interests in developing approximate, local versions of the CD Hamiltonian that retain the benefits of STA while being feasible for experimental realization~\cite{delCampoPRL2013, DeffnerPRX2014, Sels2017, Takahashi2017, Opatrny2014, BoyersPRA2019, ClaeysPRL2019, Hegade2022L, KeeverPRXQuantum2024, CepaitePRXQuantum2023, FunoNoriPRL2021, KCaiPRA2024}. 

Here we present a local counterdiabatic (CD) driving scheme for JC lattices with one excitation. Leveraging the underlying symmetries of the eigenstates in JC lattices under both periodic and open boundary conditions, we derive a local CD Hamiltonian that generates the same dynamics as the exact CD Hamiltonian. We conduct numerical simulations to demonstrate the effectiveness of this scheme in comparison with the exact CD Hamiltonian. We also utilize this method for the generation of multipartite W-states in qubits. The implementation and decoherence of this scheme in superconducting quantum devices are also discussed. This approach offers an experimentally feasible pathway to manipulating quantum states in many-body systems and can lead to further studies of the STA approach in quantum information processing.

\begin{figure}[t]
\includegraphics[clip, width=8.5cm]{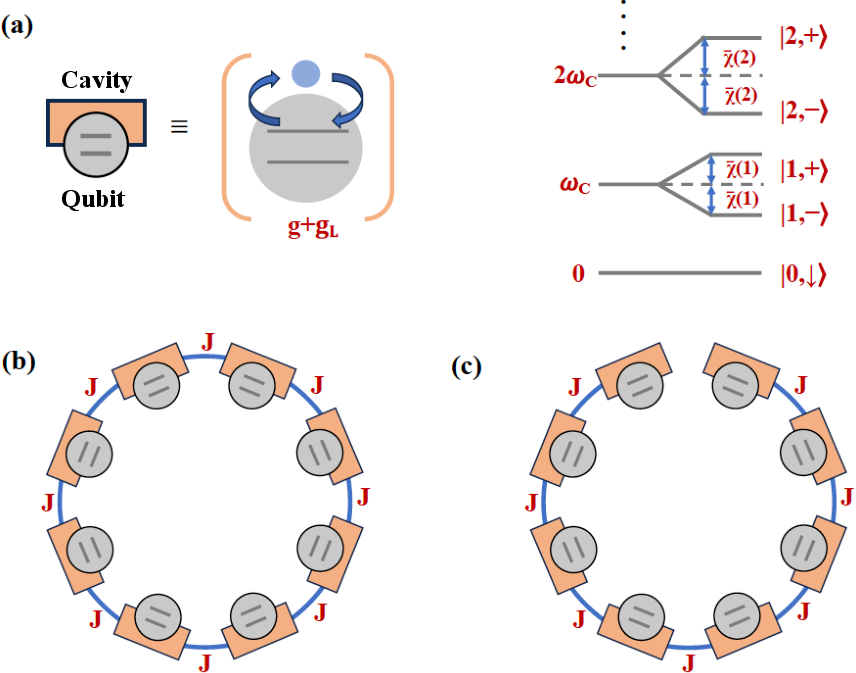}
\caption{(a) Schematic of a JC model  (left) and its eigenstates (right). Here $g$ is the qubit-cavity coupling strength, $g_{\rm L} $ is the strength of the local CD driving, and $\bar{\chi}(n)$ is the energy splitting of the eigenstates with $n$ excitations. (b) and (c) JC lattices under periodic and open boundary conditions respectively, with photon hopping rate $J$.}
\label{fig1}
\end{figure}

\emph{System and eigenstates.} A JC model contains a qubit coupled to a cavity mode with coupling strength $g$, and its eigenstates are doublets for a given number of excitations, as shown in Fig.~\ref{fig1}(a)~\cite{Larson2022Review, BlaisRMP2021cQED}. 
We consider one-dimensional JC lattices in periodic and open boundary conditions~\cite{KCaiNpj2021}, where adjacent cavity modes are connected by photon hopping with hopping rate $J$, as illustrated in Fig.~\ref{fig1}(b) and (c). The Hamiltonian of these JC lattices (assuming $\hbar=1$) can be written as $H_{\rm r} = \sum_j \Delta a_{j}^\dagger a_{j}   + g V_{g} + J V_{J}$, where $V_g= \sum_{j} ( a_{j}^{\dagger}\sigma_{j-}+\sigma_{j+}a_j)$ describes the onsite JC coupling between the qubits and the cavity modes, and $V_J = - \sum_{j} (a_{i}^{\dagger}a_{i+1}+a_{i+1}^{\dagger}a_{i})$ describes the photon hopping between neighboring sites. This Hamiltonian is written in the rotating frame of the Hamiltonian $H_{0}^{\rm (rot)}=\omega_{z}\sum_{j}(a_{j}^{\dagger}a_{j} + \frac{1}{2}\sigma_{jz})$ with $\Delta = \omega_c - \omega_z$ the detuning between the cavities and the qubits. 
Here $\omega_c$ is the cavity frequency, $\omega_z$ is the qubit energy splitting, $\sigma_{jz}$ and $\sigma_{j\pm}$ are the Pauli operators of the qubits, and $a_j$ ($a_j^\dagger$) is the annihilation (creation) operator of the cavity mode. For periodic boundary condition (PBC), the summation in $V_{J}$ is for $j\in [1,N]$ with $a_{N+1} \equiv a_{1} $, where  $N$ is the number of sites in the lattice. For open boundary condition (OBC), the summation is for $j\in [1,N-1]$. 
JC lattices have been widely studied for novel physical phenomena such as the quantum phase transition at integer fillings and the dissipative phase transition in driven systems and have been explored in various experimental systems~\cite{BWLi2022Ion, HouckPRX2017, Sala2015Nanophotonics}. 

For the JC model on site $j$, we denote the basis states by $\vert n, g\rangle_j$ and $\vert n, e\rangle_j $, where $n\ge 0$ is the photon number of the cavity and $g$ and $e$ are the qubit ground and excited states respectively.   
With a single excitation in the JC lattice, the Hilbert space is $2N$-dimensional with the basis set $\{\vert 1,g\rangle_j \prod_{j'\ne j} |0,g\rangle_{j'},\, \vert 0,e\rangle_j \prod_{j'\ne j} |0,g\rangle_{j'} \}$ and $j\in[1,N]$, which corresponds to the excitation being either a cavity photon or a qubit spin flip on site $j$. Below we will refer to this basis set as the \emph{real-space basis}, where the excitation occupies a local state. 
Under this basis, the Hamiltonian $H_{\rm r} $ can be written as a $2N\times 2N$ matrix:
\begin{equation}
    H_{\rm r}= 
    \begin{pmatrix}
     H_{\rm A} & H_{\rm B} & 0 & ... & 0 & b H_{\rm B} \\
        H_{\rm B} & H_{\rm A} & H_{\rm B}  & 0 & ... & 0 \\
        ... & ... & ... & ... & ... & ...\\
        ... & ... & ... & H_{\rm B} & H_{\rm A} & H_{\rm B} \\
        b H_{\rm B} & 0 & ... & 0 & H_{\rm B} & H_{\rm A} 
    \end{pmatrix},
    \label{eq:Hr}
\end{equation}
where ``$0$'' symbols a $2 \times 2$ zero matrix, 
\begin{equation}
    H_{\rm A} = 
    \begin{pmatrix}
        \Delta & g \\
        g & 0 
    \end{pmatrix},\quad 
     H_{\rm B} = 
    \begin{pmatrix}
        -J & 0 \\
        0 & 0 
    \end{pmatrix}, \nonumber
\end{equation}
and $b$ is the boundary condition index with $b = 1$ for PBC and $b=0 $ for OBC. For PBC, $H_{\rm r}$ is a block circulant matrix; and for OBC, $H_{\rm r}$ is a block tridiagonal Toeplitz matrix. Solving Eq.~(\ref{eq:Hr}), we find that the eigenstates of this system can be written as $\vec{w}_{k,\pm} = \vec{R}_k\otimes\vec{v}_{k,\pm}$ in the real-space basis with
\begin{equation}
        \vec{v}_{k,+} = 
        \begin{pmatrix}
            \sqrt{\frac{\chi_k+\Delta_k}{2\chi_k}} \\
            \sqrt{\frac{\chi_k-\Delta_k}{2\chi_k}}
        \end{pmatrix}, \quad
        \vec{v}_{k,-} = 
        \begin{pmatrix}
            -\sqrt{\frac{\chi_k-\Delta_k}{2\chi_k}} \\
            \sqrt{\frac{\chi_k+\Delta_k}{2\chi_k}}
        \end{pmatrix}, \nonumber
\end{equation}
$\vec{R}_k $ a N-dimensional vector, $\chi_k = \sqrt{\Delta_{k}^2+4g^2}$, and the wave vector $k\in [0, N-1]$. For PBC, the $j$th element of $\vec{R}_k $ is $(\vec{R}_k)_j = \frac{1}{\sqrt{N}}e^{i\frac{2\pi k(j-1)}{N}}$ and $\Delta_k = \Delta - 2J\cos{\frac{2 \pi k}{N}}$. 
For OBC, $(\vec{R}_k)_j  = \sqrt{\frac{2}{N+1}} \sin \frac{\pi (k+1)j}{N+1}$ and $\Delta_k = \Delta - 2J\cos{\frac{\pi (k+1)}{N+1}}$. 
For both PBC and OBC, the eigenenergies have the form $\lambda_{k,\pm} = \frac{1}{2}(\Delta_k \pm \chi_k)$ with their corresponding $\Delta_k$, and
the wave vector of the ground state is $\bar{k}=0$. Details of the derivation can be found in the Supplementary Information~\cite{SI}. We will refer to the basis set formed by the eigenstates as the \emph{k-space basis}. 

\emph{Counterdiabatic driving.} During an adiabatic evolution, a counterdiabatic (CD) Hamiltonian can be added to the system to cancel unwanted diabatic transitions. The general form of the CD Hamiltonian is~\cite{Berry2009, Demirplak2008}
\begin{equation}
    H_{\rm CD} = i\sum_{(k,l)\neq(k',l')} \frac{\vec{w}_{k,l} \vec{w}_{k,l}^\dagger \partial_t H_{\rm r} \vec{w}_{k',l'} \vec{w}_{k',l'}^\dagger}{\lambda_{k',l'}-\lambda_{k,l}}
    \label{eq:HCD}
\end{equation}
in terms of the instantaneous eigenstates $\vec{w}_{k,l}$ and eigenenergies $\lambda_{k,l}$ of the Hamiltonian $H_{\rm r}$ at time $t$. This exact CD Hamiltonian often contains non-local or multipartite interactions that are difficult to implement. 

We can calculate the exact CD Hamiltonian for a JC lattice with one excitation using the eigenstates derived above. For simplicity of discussion, we assume that the detuning $\Delta$ is a constant during the adiabatic evolution, while the couplings $g$ and $J$ are tuned linearly in the form of $g(t) = g_0 + t \partial_t g$ with $\partial_t g =(g_f - g_0)/T $ and $J(t) = J_0 + t \partial_t J$ with $\partial_t J =(J_f - J_0)/T $, where $g_0$ and $J_0$ are the initial parameters and $g_f$ and $J_f$ are the target parameters at the final time $T$. The time derivative of the adiabatic Hamiltonian is $\partial_t H_{\rm r} = \partial_t g V_g  + \partial_t J V_J$. The nonzero matrix elements of (\ref{eq:HCD}) in the k-space basis can be derived as $ \vec{w}_{k,+}^\dagger H_{\rm CD} \vec{w}_{k',-} =\delta_{k,k'}  g_{\rm CD}^{(k)}$ with
\begin{equation}
   g_{\rm CD}^{(k)} = 
    - i \frac{\Delta_k \partial_t g}{\chi_k^2}
    + i \frac{g \left(\Delta_k - \Delta\right) \partial_t J}{J \chi_k^2}. 
    \label{eq:gCDk}
\end{equation}
The exact CD Hamiltonian only includes nonzero matrix elements between eigenstates of the same wave vector $k$. This is because the variation of the adiabatic Hamiltonian only induces transitions between eigenstates of the same $k$ due to the symmetry of the eigenstates. 
Hence when written in the k-space basis, the exact CD Hamiltonian (\ref{eq:HCD}) is block diagonal with the block matrices $H_{\rm CD}^{(k)} = (0,  g_{\rm CD}^{(k)}; g_{\rm CD}^{(k)^\star}, 0)$ for all $k$ values. 
When the initial state is the ground state $\vec{w}_{\bar{k},-}$ with $\bar{k}=0$, the only allowed diabatic transition is to the excited state $\vec{w}_{\bar{k},+}$, and the only matrix element in $H_{\rm CD}$ that will affect the dynamics of the system is $g_{\rm CD}^{(\bar{k})}$.

To implement the exact CD Hamiltonian, we need to know its expression in the real-space basis, which connects directly to the physical operators of the qubits and cavities. In the Supplementary Information~\cite{SI}, we show that the exact CD Hamiltonian in the real-space basis contains nonzero off-diagonal matrices, which correspond to nonlocal couplings between qubits and cavities at different and distant sites. Such couplings are hard to implement in practical systems. 

\emph{Local CD Hamiltonian.} To construct a CD driving that only contains local couplings, we consider a block-diagonal Hamiltonian of the general form $ H_{\rm L} = I_N \otimes H_{\rm L0}$ in the real-space basis with $I_N$ a $N\times N$ identity matrix and $H_{\rm L0} = ( \delta, g_{\rm L}; g_{\rm L}^\star, -\delta)$ a $2\times2$ matrix, where $\delta$ is the detuning and $g_{\rm L}$ is the strength of the qubit-cavity coupling. The matrix elements of this Hamiltonian in the k-space basis (i.e., the eigenbasis) can be written as $ \vec{w}_{k,l}^\dagger H_{\rm L} \vec{w}_{k',l'} =\delta_{k,k'} G_{l,l'}^{(k)}$, where $G^{(k)}$ is a $2\times2$ matrix with the elements
\begin{subequations}
\begin{align}
    G_{+,+}^{(k)} & = \delta \cos (2\theta_k) +{\rm Re} [g_{\rm L}]\sin (2\theta_k), \\
    G_{+,-}^{(k)} & = - \delta \sin (2\theta_k)+{\rm Re}[g_{\rm L}]\cos (2\theta_k) +i {\rm Im}[g_{\rm L}], \\
    G_{-,+}^{(k)} & = - \delta \sin (2\theta_k)  +{\rm Re}[g_{\rm L}]\cos (2\theta_k) - i {\rm Im}[g_{\rm L}], \\
    G_{-,-}^{(k)} & = - \delta \cos (2\theta_k) - {\rm Re}[g_{\rm L}]\sin (2\theta_k),
\end{align}
\label{eq:HLpmk}
\end{subequations}
and the angle $\theta_k$ is defined by the relations $\cos \theta_k = \sqrt{\frac{\chi_k + \Delta_k}{2\chi_k}}$ and $\sin \theta_k =\sqrt{\frac{\chi_k - \Delta_k}{2\chi_k}}$.  
Hence the Hamiltonian $H_{\rm L}$ is also block diagonal in the k-space with the block matrices $G^{(k)}$.   

When the initial state is the ground state $\vec{w}_{\bar{k},-}$, the only block matrix in the exact CD Hamiltonian $H_{\rm CD}$ in the k-space basis that will affect the dynamics of the system is the matrix $H_{\rm CD}^{(\bar{k})}$. Hence, if we can make $G^{(\bar{k})} = H_{\rm CD}^{(\bar{k})}$, the dynamics under the local CD driving $H_{\rm L}$ will be exactly the same as that under the exact CD driving. For this purpose, we let $\delta=0$ and $g_{\rm L} = g_{\rm CD}^{(\bar{k})}$ with ${\rm Re}[g_{\rm L}]=0$. With (\ref{eq:HLpmk}), we find that for all $k$ values, $G^{(k)} = (0,  g_{\rm CD}^{(\bar{k})}; g_{\rm CD}^{(\bar{k})\star}, 0 )= H_{\rm CD}^{(\bar{k})}$. The local CD Hamiltonian in the k-space basis can then be written as $H_{\rm L} = I_N \otimes H_{\rm CD}^{(\bar{k})}$, which will generate the same dynamics as the exact CD Hamiltonian. An interesting observation is that $H_{\rm L}$ written in the k-space basis has exactly the same form as written in the real-space basis.  

\begin{figure}[t]
\includegraphics[clip, width=8.5cm]{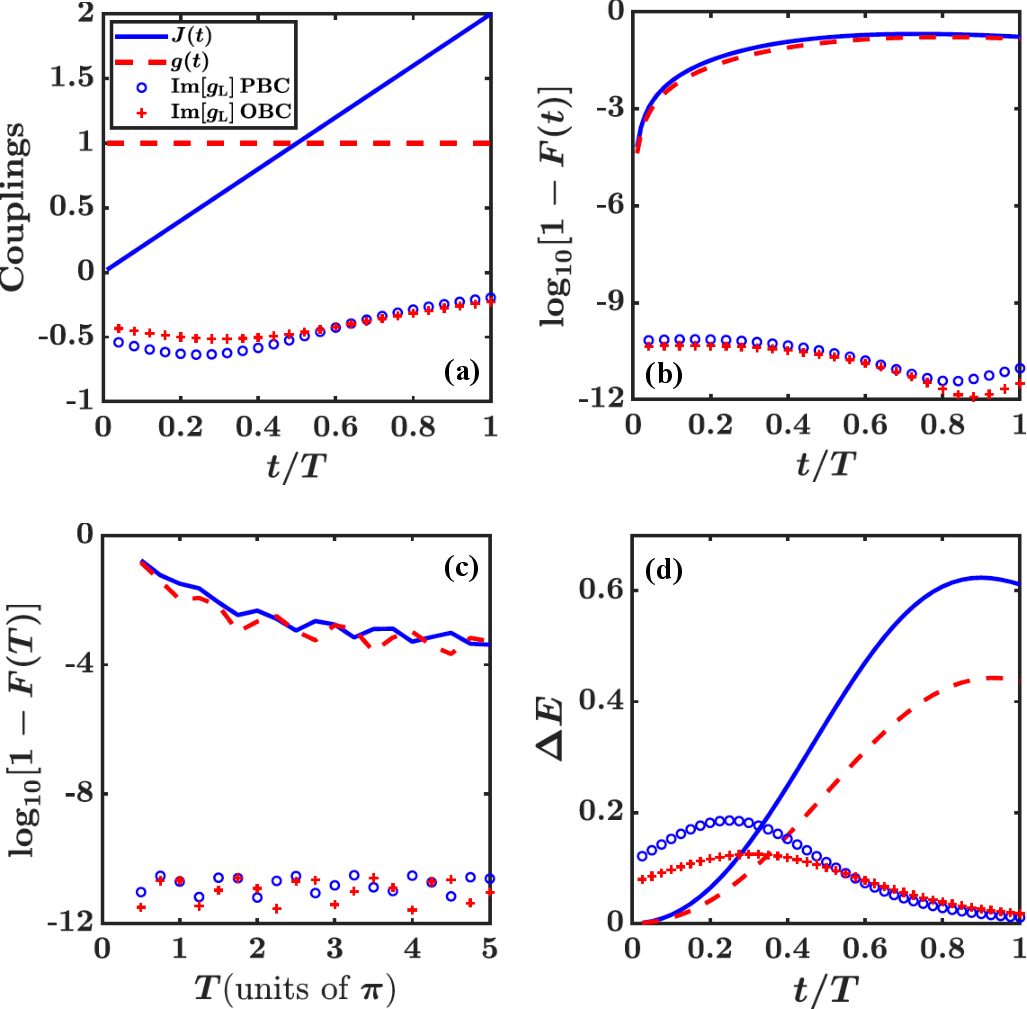}
\caption{(a) Couplings $J(t)$, $g(t)$, and ${\rm Im}[g_{\rm L}]$ for $T=0.5 \pi$. 
(b) Infidelity $1-F(t)$ vs $t/T$ for $T=0.5 \pi$. 
(c) Infidelity $1-F(T)$ vs $T$. 
(d) Energy cost $\Delta E$ vs $t/T$ for $T=0.5\pi$. 
In (b-d), the solid (dashed) line is for PBC (OBC) under $H_{\rm r}$ only; and circles (crosses) are for PBC (OBC) when $H_{\rm L}$ is applied. All parameters are in dimensionless units. }
\label{fig2}
\end{figure}
To confirm these analytical results, we conducted numerical simulations on a four-site JC lattice under both PBC and OBC. Consider an adiabatic evolution, where the photon hopping is tuned linearly from zero to the target value with $g\equiv1$, $\Delta\equiv 1$, $J(t) = J_0 + \frac{t }{T}(J_f - J_0) $, $J_0 =0$ and $J_f=2$, and the initial state is the ground state of the initial parameters.
We simulate the dynamics of this system under the adiabatic Hamiltonian $H_{\rm r}$, under the total Hamiltonian $H_t = H_{\rm r} +H_{\rm CD}$, and under $H_t = H_{\rm r} +H_{\rm L}$, respectively. 
In Fig.~\ref{fig2}(a), the couplings $g(t)$ and $J(t)$ are plotted together with the local CD coupling ${\rm Im}[g_{\rm CD}]$. It shows that the local CD coupling is of the same order of magnitude as the couplings $g$ and $J$. 
The fidelity of the system's state at time $t$ is defined as $F(t) = \vert\vec{w}^\dagger(t) \vec{w}_{\bar{k},-}(t)\vert^2$ with regard to the instantaneous ground state $\vec{w}_{\bar{k},-}(t)$ of $H_{\rm r}$.  
To characterize the performance of the scheme, we plot the infidelity $1-F(t)$ vs time $t$ in Fig.~\ref{fig2}(b). The infidelity under the local CD Hamiltonian remains negligibly small during the entire evolution, demonstrating the effectiveness of the local CD driving.  The infidelity $1-F(T)$ vs the total time $T$ in Fig.~\ref{fig2}(c) also confirms this observation. The numerical results of the infidelity under the exact CD Hamiltonian $H_{\rm CD}$ are identical to that of $H_{\rm L}$ up to a numerical error below $10^{-12}$, as shown in the Supplementary Information~\cite{SI}. 

Another important quantity for the CD driving is the energy cost defined by $\Delta E = \vec{w}^\dagger(t) H_{\rm t} \vec{w}(t) - E_{G}$, which is the difference between the average energy of the system and the instantaneous ground state energy $E_{G}$ of the total Hamiltonian $H_{\rm t}$ at time $t$~\cite{AbahPRE2019}. 
In Fig.~\ref{fig2}(d), we plot $\Delta E$ vs time $t$. At $t=0$, $\Delta E=0$ under the adiabatic Hamiltonian $H_{\rm r}$ because the initial state is the ground state of $H_{\rm r}$. The energy difference increases with time due to the diabatic transitions to the excited states. When the local CD Hamiltonian is applied, $\Delta E\ne 0$ at $t=0$, because the ground state of $H_{\rm t}$ is not the ground state of $H_{\rm r}$. During the evolution, $\Delta E$ does not increase significantly as the diabatic transitions are eliminated by the CD driving. 

\begin{figure}[t]
\includegraphics[clip, width=8.5cm]{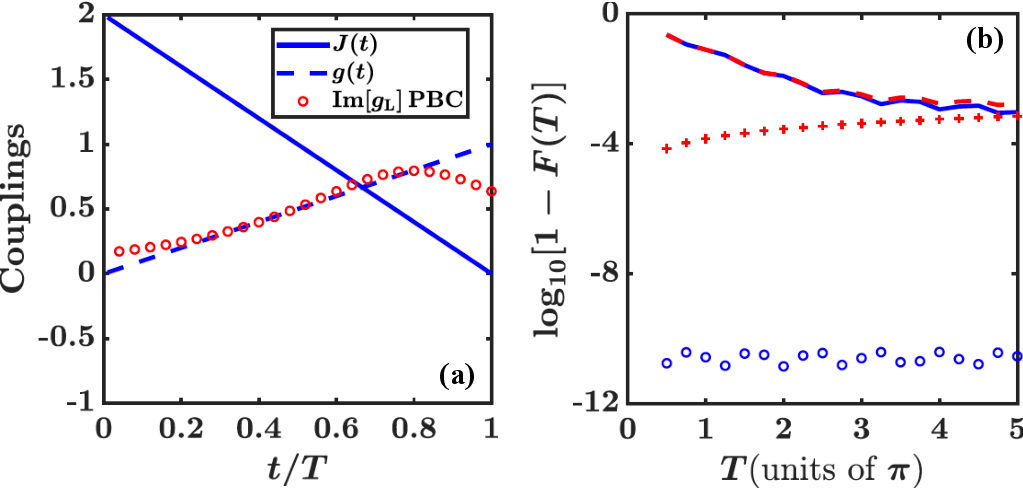}
\caption{W-state generation. (a) Couplings $J(t)$, $g(t)$, and ${\rm Im}[g_{\rm L}]$ for $T=0.5 \pi$.
(b) Infidelity $1-F(T)$ vs $T$. The solid (dashed) line is for $\kappa=\gamma=0$ ($\kappa=5\times 10^{-5}$ and $\gamma=\frac{5}{\pi}\times 10^{-5}$) under $H_{\rm r}$ only; and circles (crosses) are for $\kappa=\gamma=0$ ($\kappa=5\times 10^{-5}$ and $\gamma=\frac{5}{\pi}\times 10^{-5}$) when $H_{\rm L}$ is applied. All parameters are in dimensionless units.}
\label{fig3}
\end{figure}
\emph{W-state in JC lattice.}  
The W-state is a prototype multipartite entangled state that preserves the entanglement at the loss of some qubits (or subsystems)~\cite{Wstate}. For a $N$-qubit system, the W-state has the form $ \vert W\rangle  = \frac{1}{\sqrt{N}}\sum_{j=1}^{N} \vert e\rangle_j \prod_{j'\ne j} \vert g\rangle_{j'} $. Leveraging the local CD driving scheme, we show that in a JC lattice with one excitation, this state can be prepared with high fidelity in timescales much shorter than the decoherence times. 
Consider an adiabatic process with linear ramping of the couplings $J$ and $g$, $\Delta\equiv 1$, $g_0=0$, $g_f=1$, $J_0=2$, and $J_f=0$ under PBC. At $t=0$, the system is prepared in the ground state of the initial parameters, which can be written in vector form as $w(0) = \vec{R}_{\bar{k}} \otimes\vec{v}_{\bar{k},-} $ with $(\vec{R}_{\bar{k}})_j =1/\sqrt{N}$ for $\bar{k}=0$ and $\vec{v}_{\bar{k},-}=(-1,0)^T$ (here $\chi_k = \vert\Delta_k\vert$ since $\Delta_k<0$). This state is an equal superposition of the $n=1$ photon Fock states of all cavity modes. At $t=T$, the final state also has the form $w(T) = \vec{R}_{\bar{k}}\otimes\vec{v}_{\bar{k},-} $, but with $g_f=1$ and $J_f=0$, $\vec{v}_{\bar{k},-}$ becomes the lower eigenstate of a single JC model with one excitation $\vert 1,-\rangle$. With $J_f=0$, the JC models in the lattice are isolated from each other. The state $\vert 1,-\rangle_j$ can then be rotated to the state $\vert 0,e\rangle_j$ in the JC models by local operations, and the $N$-qubit W-state can be generated. 
We numerically simulate this process for a four-site JC lattice under PBC. The couplings $g(t)$, $J(t)$ and the local CD driving strength are plotted in Fig.~\ref{fig3}(a) and the infidelity vs the total time $T$ is plotted in Fig.~\ref{fig3}(b) both without and with the decoherence terms. It is shown that at a short evolution time $T=0.5\pi$, the infidelity of the generated state can be below $10^{-4}$ even in the presence of decoherence.

\begin{figure}[t]
\includegraphics[clip, width=8cm]{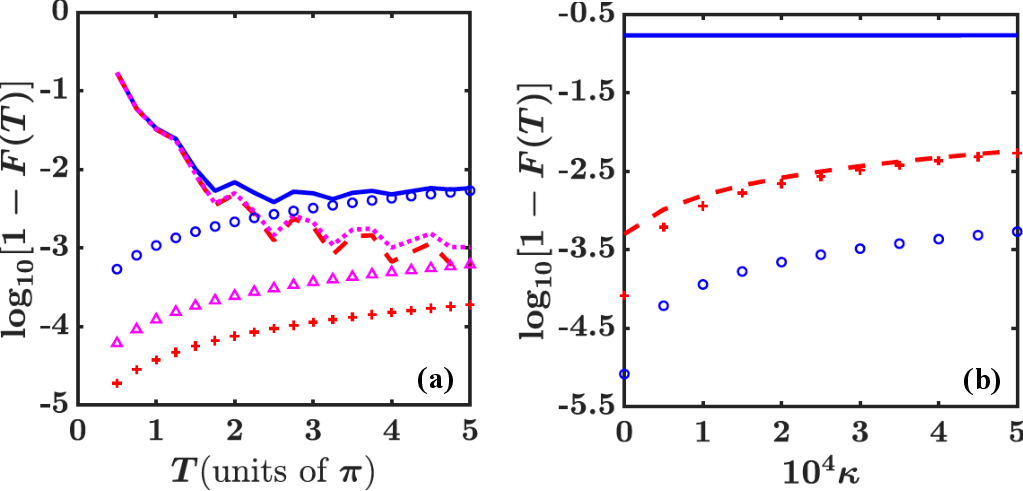}
\caption{Decoherence. (a) Infidelity $1-F(T)$ vs $T$. The solid, dotted, and dashed lines are for $\kappa=5\times 10^{-4}$, $\kappa=10^{-4}$, and $\kappa=5\times 10^{-5}$ under $H_{\rm r}$ only; and the circles, triangles, and crosses are for these values of $\kappa$ when $H_{\rm L}$ is applied. (b) Infidelity $1-F(T)$ vs $\kappa$. The solid (dashed) line is for $T=0.5\pi$ ($T=5\pi$) under $H_{\rm r}$ only; and the circles (crosses) are for $T=0.5\pi$ ($T=5\pi$) when $H_{\rm L}$ is applied. Here $\gamma=\frac{5}{\pi}\times 10^{-5}$, and all parameters are in dimensionless units.}
\label{fig4}
\end{figure}
\emph{Physical implementation.} JC lattices have been realized in various physical platforms, including superconducting qubits coupled to resonators, defects or quantum dots coupled to nanocavities, internal states coupled to the motional states of trapped ions~\cite{Noh2017Review, 2012HouckNP_JCQS, BWLi2022Ion, HouckPRX2017, Sala2015Nanophotonics}. We take the superconducting system as an example, where many forms of couplings have been demonstrated in experiments~\cite{SiddiqiPRL2021, CampbellLaHayePRApplied2023, Chen:2014, FYanPRApplied2018, SandbergAPL2008}. 
The parameters in our numerical simulations are based on practical parameters.   
We assume that the dimensionless coupling strength $g=1$ corresponds to $2\pi\times 100$ MHz. The dimensionless time $T=0.5\pi$ ($T=5\pi$) then corresponds to $2.5$ ($25$) ns. The magnitude of the local CD driving depends on the time derivative of the coupling(s). Using (\ref{eq:gCDk}), we find that this magnitude should be smaller than $\vert \partial_t g\vert  / \chi_{\bar{k}}$ or $\vert \partial_t J \vert/2 \chi_{\bar{k}}$. As shown in Figs.~\ref{fig2} and \ref{fig3}, this magnitude is on the same order of magnitude as the couplings.
A typical qubit decoherence time of $100 {\rm \mu s}$ corresponds to a dimensionless decoherence rate $\gamma=\frac{5}{\pi}\times 10^{-5}$. For a cavity frequency of $\omega_c/2\pi\sim 5$ GHz and a quality factor $Q=10^6$, the cavity decay rate is $\kappa=5\times 10^{-5}$ in dimensionless unit. 
We numerically simulate the effect of decoherence using the master equation $d_t\rho = -i\left [H_{\rm t}, \rho\right] + \sum_j \left(\gamma_j \mathcal{L}_{qj}+ \kappa_j\mathcal{L}_{aj}\right) \rho$, where $\mathcal{L}_{qj} \rho = \frac{1}{2}(2\sigma_{j-}\rho\sigma_{j+}- \rho\sigma_{j+}\sigma_{j-}-\sigma_{j+}\sigma_{j-}\rho)$ is the Liouvillian operator for the qubit on site $j$ with damping rate $\gamma_j$ and $\mathcal{L}_{aj}\rho =\frac{1}{2} (2a_j \rho a_j^\dagger- \rho a_j^\dagger a_j -a_j^\dagger a_j \rho)$ is the Liouvillian operator for the cavity on site $j$ with decay rate $\kappa_j$. We assume $\gamma_{j}=\gamma$ and $\kappa_{j}=\kappa$ for simplicity of discussion.
In Fig.~\ref{fig4}(a) and (b), we plot the infidelity vs the total time $T$ and the cavity decay rate $\kappa$, respectively. The numerical results clearly confirm that the large infidelity at a short time $T$ is mainly caused by diabatic transitions, which is suppressed by the local CD driving. For a long evolution time $T$, the infidelity is dominated by decoherence effects, and hence the local CD driving is not effective in this case.

\emph{Conclusions.} Leveraging the symmetries of the eigenstates in JC lattices with one polariton excitation, we found a local CD driving that only requires onsite qubit-cavity couplings. The local CD driving eliminates all diabatic transitions during an adiabatic evolution and generates the same dynamics as the exact CD driving. We conducted numerical simulation to confirm the analytical results and applied the scheme to the generation of the multipartite W-state. The implementation and decoherence of this system have also been discussed. This work can lead to further studies of the realization of STA approaches in quantum information systems. 

\emph{Acknowledgements.} This work is supported by the NSF Award No. 2037987 and the UC-MRPI Program (Grant ID M23PL5936). A.G. is also supported by the Sandbox AQ Fellowship.

\end{document}